\documentclass[11pt,twoside]{article}
\usepackage{asp2006}
\usepackage{epsf}
\begin{document}
\title{Historic mass loss from the RS\,Ophiuchi system}
\author{Jacco Th.\ van Loon}
\affil{Lennard-Jones Laboratories, Keele University, Staffordshire ST5 5BG,
United Kingdom; jacco@astro.keele.ac.uk, www.astro.keele.ac.uk/$\sim$jacco}

\begin{abstract}
Dust has been detected in the recurrent nova RS\,Ophiuchi on several
occassions. I model the historical mid-infrared photometry and a recent
Spitzer Space Telescope spectrum taken only half a year after the 2006
eruption. The dust envelope is little affected by the eruptions. I show
evidence that the eruptions and possibly the red giant wind of RS\,Oph may
sculpt the interstellar medium, and show similar evidence for the recurrent
dwarf nova T\,Pyxidis.
\end{abstract}

\section{Dusty stellar winds}

\subsection{Asymptotic Giant Branch stars and red supergiants}

Stars with an initial mass between $\sim1$ and $\sim40$ M$_\odot$ become
hydrogen/helium-shell-burning Asymptotic Giant Branch (AGB) stars or
core-helium-burning red supergiants (RSG). These phases are characterised by
cool, molecular atmospheres giving rise to M spectral types (S or C for some
chemically peculiar AGB stars), strong radial pulsations of this atmosphere on
timescales of a year or more, and a high luminosity ($L>2,000$ L$_\odot$). The
pulsation may lift the atmosphere high enough such that dust can condense
\citep{BowenWillson1991}. Radiation pressure on the grains then drives a dust
wind which, via collisions with molecular hydrogen, drags the gas along with
it. At a typical wind speed $v_\infty\sim5$ to 30 km s$^{-1}$ these stars lose
mass at rates from $\dot{M}\sim10^{-7}$ to over $10^{-4}$ M$_\odot$ yr$^{-1}$
\citep{vanLoonEtal1999}. In AGB stars this leads to the removal of the mantle
and the premature death of the star, leaving the truncated core behind as a
cooling white dwarf. The more massive RSGs either explode or evolve back to
higher surface temperatures, and it is not yet clear how much mass they will
have shed during the preceding RSG stage.

Mass-loss rates of cool, luminous stars are most easily estimated from the
reprocessed radiation emitted by the dust grains at infrared (IR) wavelengths,
which is particularly conspicuous in the $\lambda\sim10$ to 40 $\mu$m region.
The main problem with this method is that the dust constitutes only a minor
fraction of the total mass in the wind, typically $\sim1$:$200$
\citep{Knapp1985} but poorly known for all but the dustiest stars. Carbon
monoxide has strong rotational transitions at mm wavelengths which can be
detected in relatively nearby stars, but despite being the most abundant
molecule after H$_2$ this too is a trace species of which the mass fraction is
uncertain --- and modelling the CO line requires knowledge of the temperature
profile throughout the envelope which depends, amongst other things, on the
dust content.

\subsection{First ascent Red Giant Branch stars}

Low-mass stars ($M_{\rm initial}<2$ M$_\odot$) evolve along a first ascent Red
Giant Branch (RGB) as hydrogen-shell burning stars with an inert helium core.
They reach a maximum luminosity of only $L_{\rm tip}\sim2,000$ L$_\odot$, and
it becomes problematic for them to drive a wind. The situation is worsened by
the fact that many RGB stars do not pulsate strongly and slowly enough to
sufficiently increase the scaleheight, and the dust condensation is therefore
unlikely to reach completeness. With a low, uncertain dust:gas mass ratio and
possibly higher dilution of the dust envelope as it kinematically decouples
from the bulk gas, measured mass-loss rates will tend to be too low. The
threshold below which this happens is not well known, partly because of
uncertainties in the opacity of the grains as they nucleate and grow:
\citet{GailSedlmayr1987} estimate $\dot{M}\gg 10^{-6}$ M$_\odot$ yr$^{-1}$ but
\citet{NetzerElitzur1993} place it at $\dot{M}>10^{-7}$ M$_\odot$ yr$^{-1}$.
\citet{JudgeStencel1991} show that RGB mass loss must be driven by another
mechanism, but that this appears to be similarly efficient as a dust-driven
wind. The warmer stars further down the RGB have more prominent chromospheres,
which may be accompanied by the generation of Alfv\'en or acoustic waves that
provide a possible alternative mechanism for driving a wind.

Mass-loss rates estimated from the IR emission for the brightest, dustiest RGB
stars are typically $\dot{M}\sim10^{-7}$ to $10^{-6}$ M$_\odot$
\citep{vanLoonEtal2006,OrigliaEtal2007}. Mass-loss rates from
$\dot{M}\sim10^{-9}$ to $10^{-6}$ M$_\odot$ yr$^{-1}$ have been determined
from the blue-displaced cores of strong optical absorption lines and emission
in some of these lines, but these are very uncertain too because of the
sensitivity to the precise excitation and ionization conditions in the wind.
For the most luminous RGB stars the IR and optical estimates tend to agree
within an order of magnitude \citep{McDonaldvanLoon2007}.

\begin{figure}[!t]
\plotone{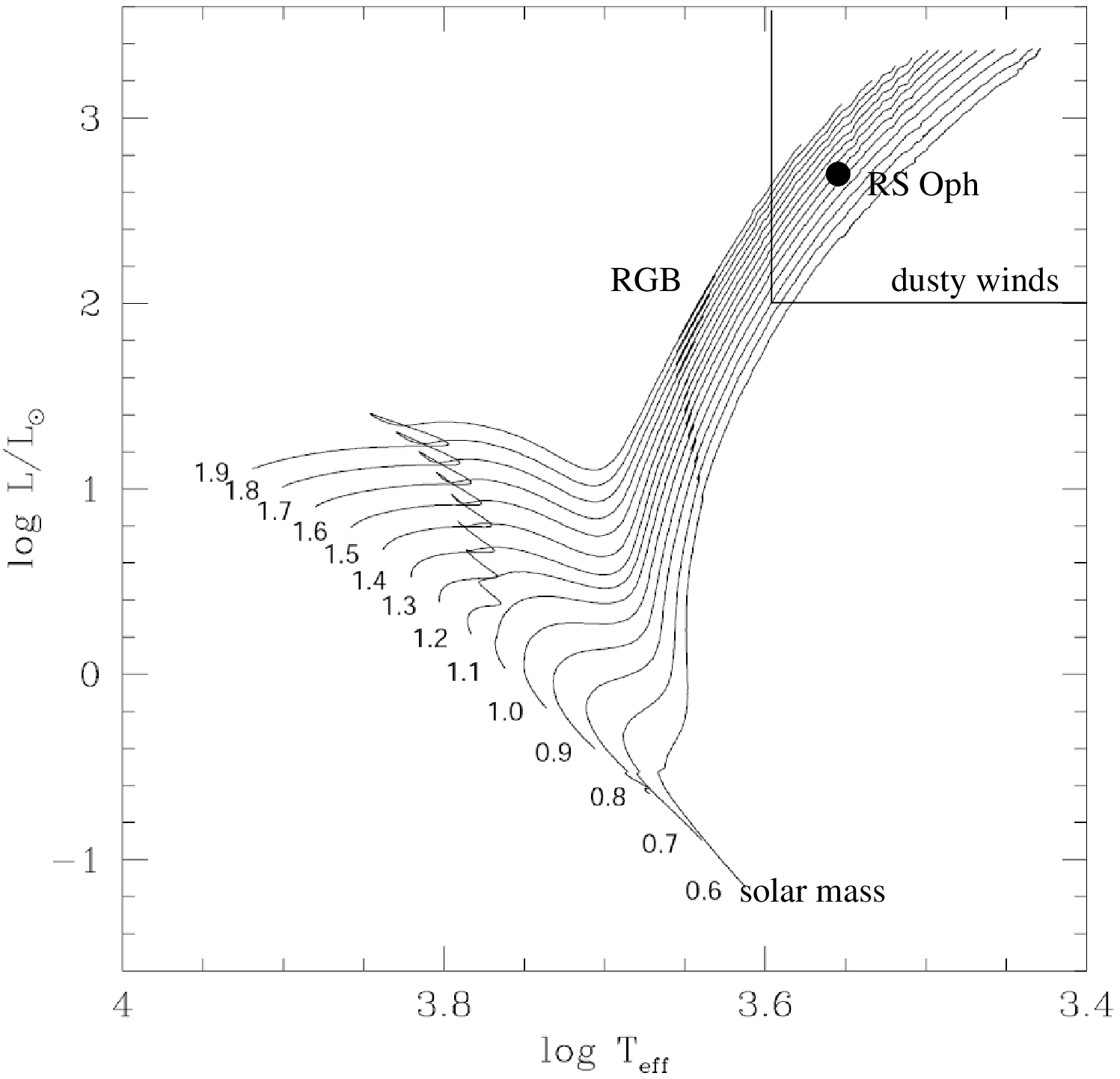}
\caption{RS\,Oph compared to isochrones from \citet{GirardiEtal2000}.}
\end{figure}

\section{The dusty stellar wind of RS\,Ophiuchi}

RS\,Ophiuchi has a spectral type of M2 \citep{FekelEtal2000}, and I will later
show that the luminosity is $L\sim444$ L$_\odot$. This places it on an
isochrone of a star with an initial mass of $M_{\rm initial}\sim1.1\pm0.1$
M$_\odot$ (Fig.\ 1). Although it is a few times fainter than the tip-RGB
luminosity and not as cool as it gets, mass loss is expected at a level of
$\dot{M}>10^{-9}$ M$_\odot$ yr$^{-1}$ and may be accompanied by dust.

\subsection{Mid-IR observations}

RS\,Oph is usually classified as an S-type system, which by definition is a
system devoid of dust as the dusty systems are classified as D-type. The
latter usually contain a Mira-type large-amplitude variable donor star, which
is likely to be an AGB star. RS\,Oph is not a Mira-type variable, and
definitely on the RGB, but it does have circumstellar dust, and the S-type
classification is thus misleading. Its IR emission at $\lambda=10$ $\mu$m was
detected and attributed to dust for the first time in 1970 by
\citet{GeiselKleinmannLow1970}, which lead them to remark: ``The infrared
emission from novae closely resembles that from many infrared stars in which
the rate of mass loss is more constant. This suggests that the formation of
grains is governed by similar processes in all these systems, and that novae
are key objects in understanding how the dust is formed.''

Dust had thus been detected within three years from the eruption in 1967.
RS\,Oph underwent two more eruptions, in 1985 and in 2006, and as far as I
know the system has been observed at mid-IR wavelengths twice more during
quiescence and again shortly after the 2006 eruption (Fig.\ 2). Large
differences in the mid-IR emission might be expected depending on the temporal
proximity of an eruption, so it came as a surprise to find the contrary.

\begin{figure}[!b]
\plotone{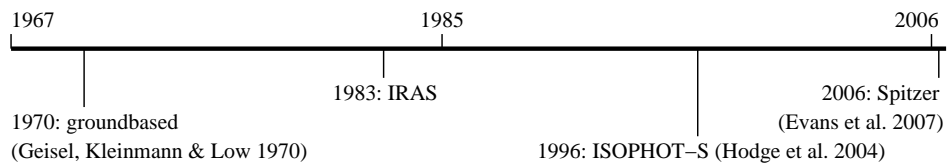}
\caption{Times of eruptions and mid-infrared observations.}
\end{figure}

RS\,Oph was detected in the InfraRed Astronomical Satellite (IRAS) survey in
1983 \citep{Schaefer1986}. It is listed in the IRAS Point Source Catalogue
with $F_{12}=0.4$ Jy and $F_{25}<0.32$ Jy (the upper limits at 60 and 100
$\mu$m are meaningless); \citet*{KenyonFernandezCastroStencel1988} estimate
$F_{12}=0.38$ Jy and $F_{25}=0.12$ Jy, and \citet{HarrisonGehrz1994} estimate
$F_{12}=0.41\pm0.04$ Jy and $F_{25}=0.14\pm0.04$ Jy. I have re-measured the
IRAS scans, and derive refined estimates of $F_{12}=0.37\pm0.02$ Jy and
$F_{25}=0.09$:$\pm0.03$ Jy (Fig.\ 3).
\citet*{KenyonFernandezCastroStencel1986} estimated $F_{12}=0.30$ Jy and
$F_{25}=0.16$ Jy from pointed observations carried out with IRAS in September
1983 (they estimate that free-free emission may contribute $\sim0.03$ and
$\sim0.04$ Jy at $\lambda=12$ and 25 $\mu$m, respectively). The IRAS 12-$\mu$m
flux densities are in remarkably good agreement with the
\citet{GeiselKleinmannLow1970} measurement of $F_{10.4}\sim0.4$ Jy (an N-band
magnitude of $m_{\rm N}\sim4.5$ to 5.1 mag) a dozen years earlier. Then again,
there had not been any eruption meanwhile.

Soon after the end of the IRAS mission, RS\,Oph put up a show again.
Curiously, it took more than a decade until new mid-IR measurements were made,
in March 1996 with the PHOT-S instrument onboard the Infrared Space
Observatory (ISO). \citet{HodgeEtal2004} analysed the spectrum shortward of
$\lambda=9$ $\mu$m and found no evidence for dust emission. However, at those
wavelengths the dust emission is $<0.1$ Jy; the uncertainties in zodiacal
light correction and instrument calibration and sensitivity do not permit the
detection of the dust emission in the PHOT-S spectrum around 8-11 $\mu$m.

\begin{figure}[!t]
\plotone{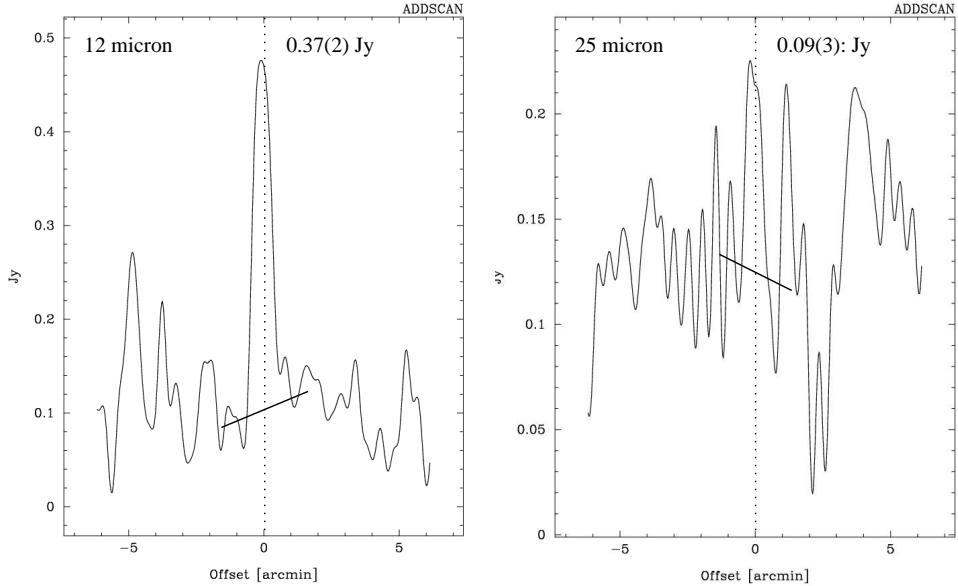}
\caption{IRAS scans through RS\,Oph at 12 $\mu$m (left) and 25 $\mu$m
(right).}
\end{figure}

Another decade passed before new mid-IR measurements were made, this time with
the Spitzer Space Telescope (Spitzer) triggered by the February 2006 eruption.
By September 2006 the free-free and line emission had weakened enough to
reveal a continuum and spectral features unambiguously identifiable with
predominantly silicate-type dust \citep{EvansEtal2007}. Remarkably, the
Spitzer spectrum agrees frightfully well with all previous photometry (Fig.\
4).

\subsection{Modelling the dust envelope}

\begin{figure}[!t]
\plotone{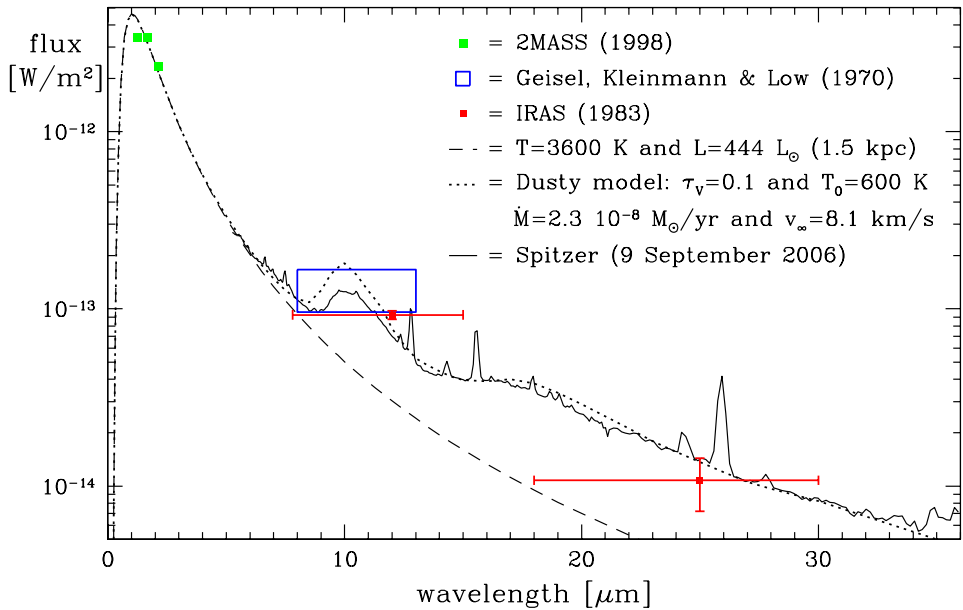}
\caption{Photometry and Spitzer spectrum, overlain with a black body
representing the star and a Dusty model that reproduces the envelope.}
\end{figure}

I used the radiative transfer code Dusty \citep{IvezicElitzur1997} to model
the spectral energy distribution (SED). The main input parameter of interest
is the optical depth, $\tau_{\rm V}$, which is related to the mass-loss rate
via the dust:gas mass ratio (assumed to be 1:200), luminosity ($L=444$
L$_\odot$, measured from the fit to the data assuming a distance of 1.5 kpc)
and wind speed (computed by the model for a dust-driven wind). The second
input parameter of interest is the temperature at the inner edge of the dust
envelope, T$_0$. The model SED also depends on the dust species used, and on
the temperature of the star (for which we adopt $T=3,600$ K consistent with
its spectral type). A good fit was obtained using circumstellar silicates from
\citet*{OssenkopfHenningMathis1992}, and $T_0=600$ K. The model overestimates
the strength of the 10-$\mu$m feature, but this also depends on the grain
properties such as size, shape and composition (a
\citep*{MathisRumplNordsieck1977} size distribution was assumed).

I thus estimate a mass-loss rate $\dot{M}=2.3\times10^{-8}$ M$_\odot$
yr$^{-1}$ at a wind speed $v_\infty=8.1$ km s$^{-1}$. Although silicates can
form already at temperatures above 1,000 K there is a conspicuous absence of
such warm dust ($T>600$ K). Because the dust shell is optically thin the SED
does not depend much on the geometry. Assuming a spherically symmetric wind,
the dust envelope starts at a distance of 39 R$_\star$ ($\sim10$ AU) from the
red giant (the outer radius is not constrained by the available data other
than that it is $\gg 10^2$ R$_\star$); the orbital radius is $\sim1$ AU (Fig.\
5). Although the wind is unlikely to travel at a constant speed $v_\infty$
before dust even forms, it would take 6 years to reach the dust formation zone
if it did; the binary would have gone round nearly five times and it is thus
highly unlikely that the binary motion leaves non-axisymmetric imprints on the
dust shell.

\begin{figure}[!t]
\plotone{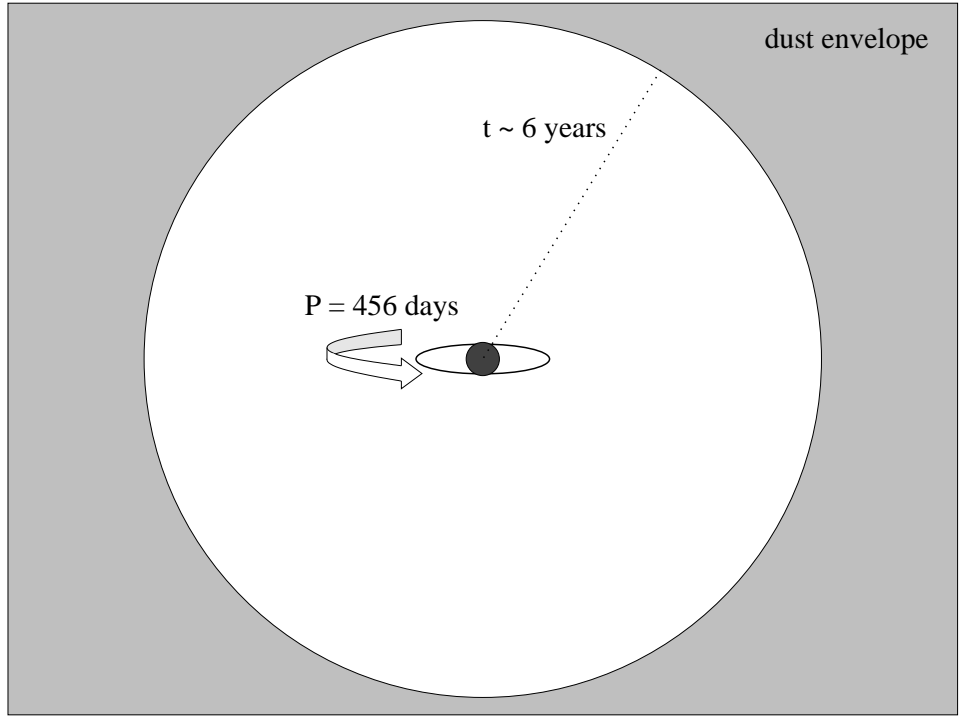}
\caption{Sketch of the geometry of the RS\,Oph system, roughly to scale.}
\end{figure}

\subsection{How typical is the mass-loss rate of RS\,Oph?}

\begin{figure}[!t]
\plotone{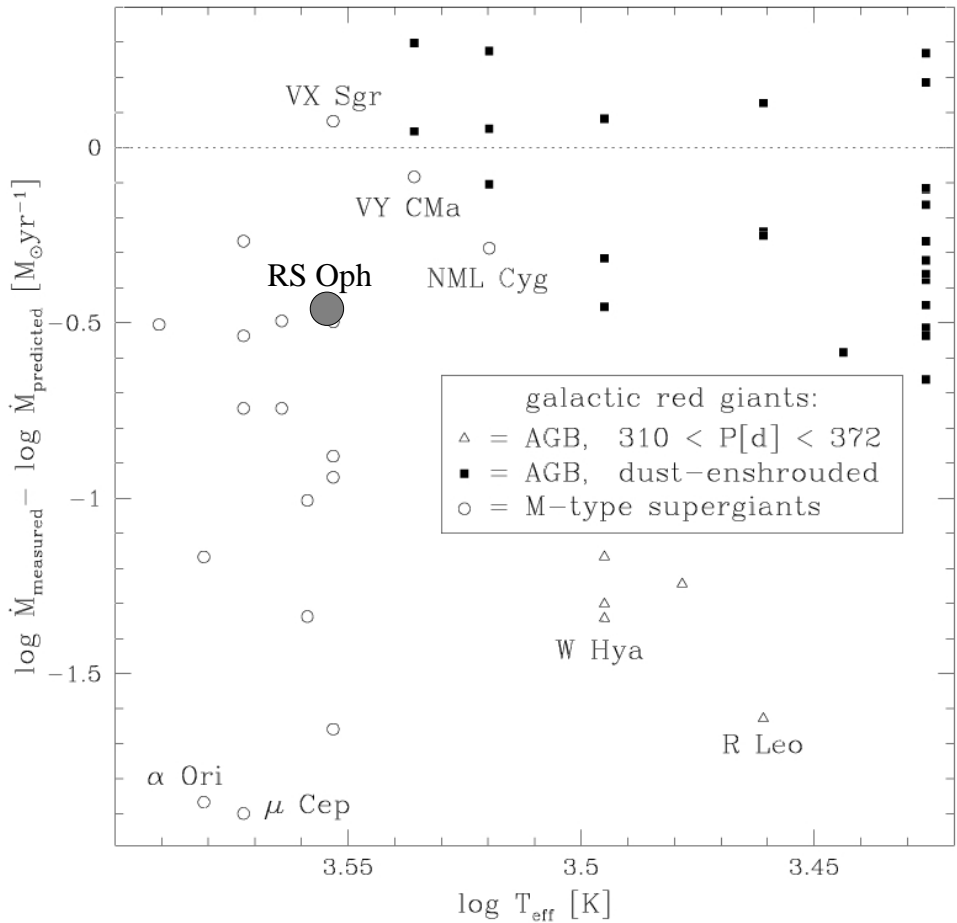}
\caption{RS\,Oph compared with other dusty red giants, and with predictions.}
\end{figure}

It has been suggested that symbiotic stars lose mass more rapidly than single
stars \citep{Kenyon1988}. However, when comparing RS\,Oph with other dusty
stars I find no evidence for a particularly high mass-loss rate. In Fig.\ 6
the differences are plotted between the measured mass-loss rate and the one
predicted by the formula for dust-driven winds of oxygen-rich AGB stars and
RSGs \citep{vanLoonEtal2005}. RS\,Oph sits at the warm end of the diagram,
where many stars appear to have lower mass-loss rates or a lower dust content
(including the famous RSGs Betelgeuse and $\mu$\,Cep). Yet its mass-loss rate
is nearly as expected for a well-developed, efficiently dust-driven wind.
Although at the faint end of the distribution, RS\,Oph is also in line with
what is measured for red giants in globular clusters \citep[their Fig.\
13]{McDonaldvanLoon2007}.

On the other hand, stars that do not pulsate in the fundamental mode, like
Miras do, have lower mass-loss rates or less complete dust formation (Fig.\
6). Compared to those cooler stars, RS\,Oph seems to have a rather high
mass-loss rate. The mass-loss rate of RS\,Oph may even have been
under-estimated, if the dust:gas ratio is lower than assumed. Physical
mechanisms exist that might explain enhanced mass loss in symbiotic systems,
e.g., through equatorial collimation or tides \citep{Mikolajewska2000}.

The remarkable observation is that the system has been losing mass at a (near)
constant rate, and that the eruptions have done little to affect the dust
envelope. This can be understood if the eruptions are highly an-isotropic.

\section{Do recurrent novae affect the interstellar medium?}

Recurrent novae are relatively old systems, and hence they have developed
large peculiar velocities with respect to the ISM. It is therefore possible
that recurrent novae carve out a path through the ISM as their winds sweep up
ISM material and/or the nova eruptions leave a trail of successive shells.

\subsection{The case of RS\,Oph}

A composite of IRAS images at 12, 25 and 60 $\mu$m reveals a dark patch of
1-2$^\circ$ diameter around RS\,Oph (Fig.\ 7) --- the direction of the
galactic plane is towards the lower-left corner. After scaling to the same
mean and spread, a median combination of all 12-100 $\mu$m images shows an
elongated cut into the ISM, with RS\,Oph near its end and its proper motion
pointing at a trajectory along it. In the same time it took RS\,Oph to
traverse along the structure ($\sim1$ Myr), its wind would have blown a bubble
of $\sim1^\circ$ diameter (Fig.\ 7). This is highly suggestive of RS\,Oph
being responsible for the ISM structure.

\begin{figure}[!t]
\plotone{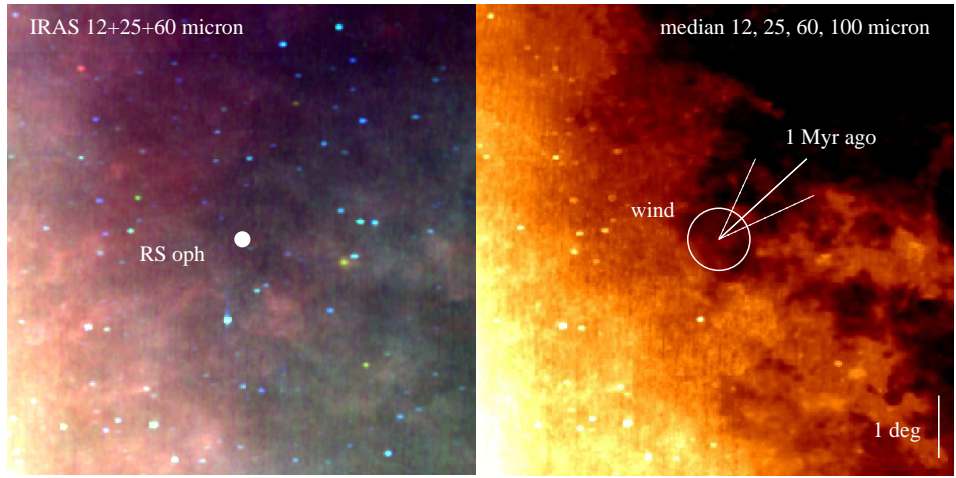}
\caption{IRAS composites around RS\,Oph (North is up, East is left); the right
panel is a median combination of scaled 12, 25, 60 and 100 $\mu$m images.}
\end{figure}

\begin{figure}[!t]
\plotone{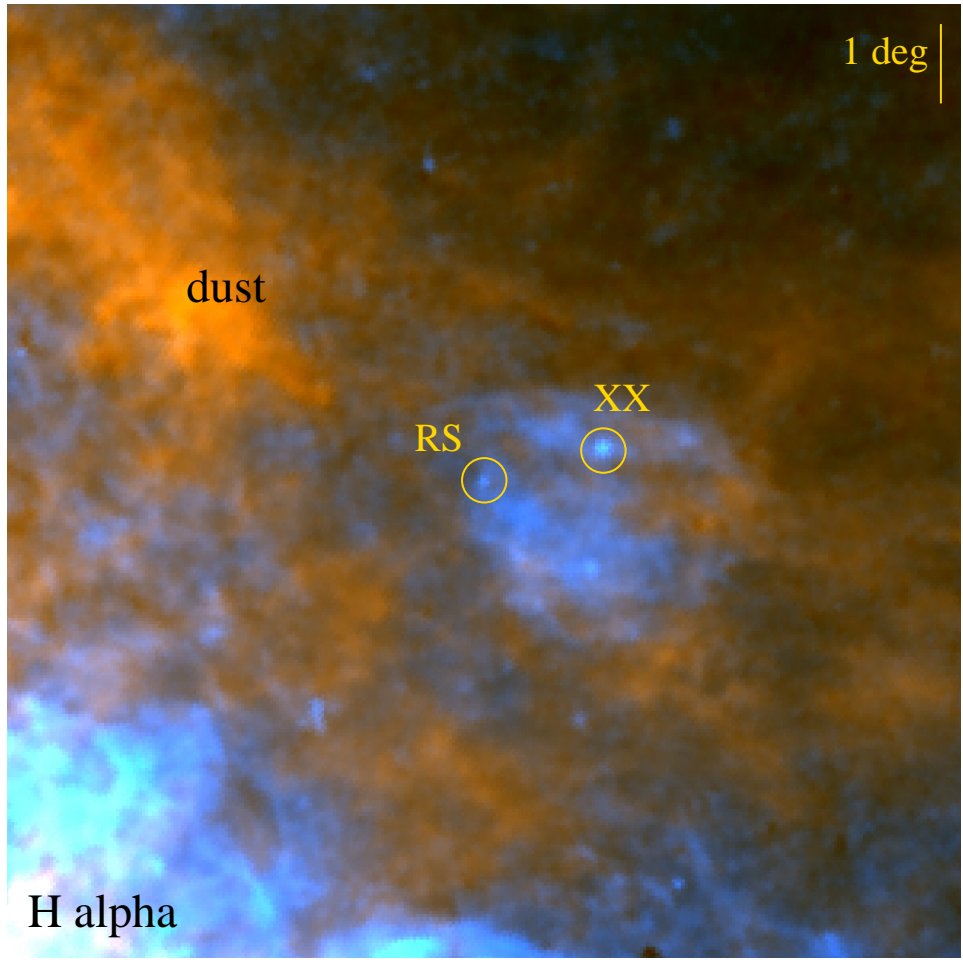}
\caption{Combination of the COBE dust map (orange) and SHASSA H$\alpha$ survey
(blue), showing RS\,Oph and the nearby symbiotic binary XX\,Oph.}
\end{figure}

An H$\alpha$ map \citep{Finkbeiner2003} shows what looks like a bubble of
ionized gas, with RS\,Oph near its western edge (Fig.\ 8).
\citet{CoolEtal2005} noticed this structure when studying the nearby symbiotic
binary XX\,Oph; they estimate a radial velocity of $v=2.4\pm11.8$ km s$^{-1}$
for the gas and exclude an association with XX\,Oph. RS\,Oph has $v=-40$ km
s$^{-1}$ \citep{FekelEtal2000} and is therefore also not associated with the
extended H$\alpha$ emission. However, combined with the dust map from
\citet*{SchlegelFinkbeinerDavis1998} the ISM structure more closely resembles
a hole in the dust through which background emission is seen, rather than a
hot bubble (Fig.\ 8). Composed of a Be star and a late-M giant, XX\,Oph rather
than RS\,Oph might have created the ISM structure.

\subsection{The case of T\,Pyx}

\begin{figure}[!t]
\plotone{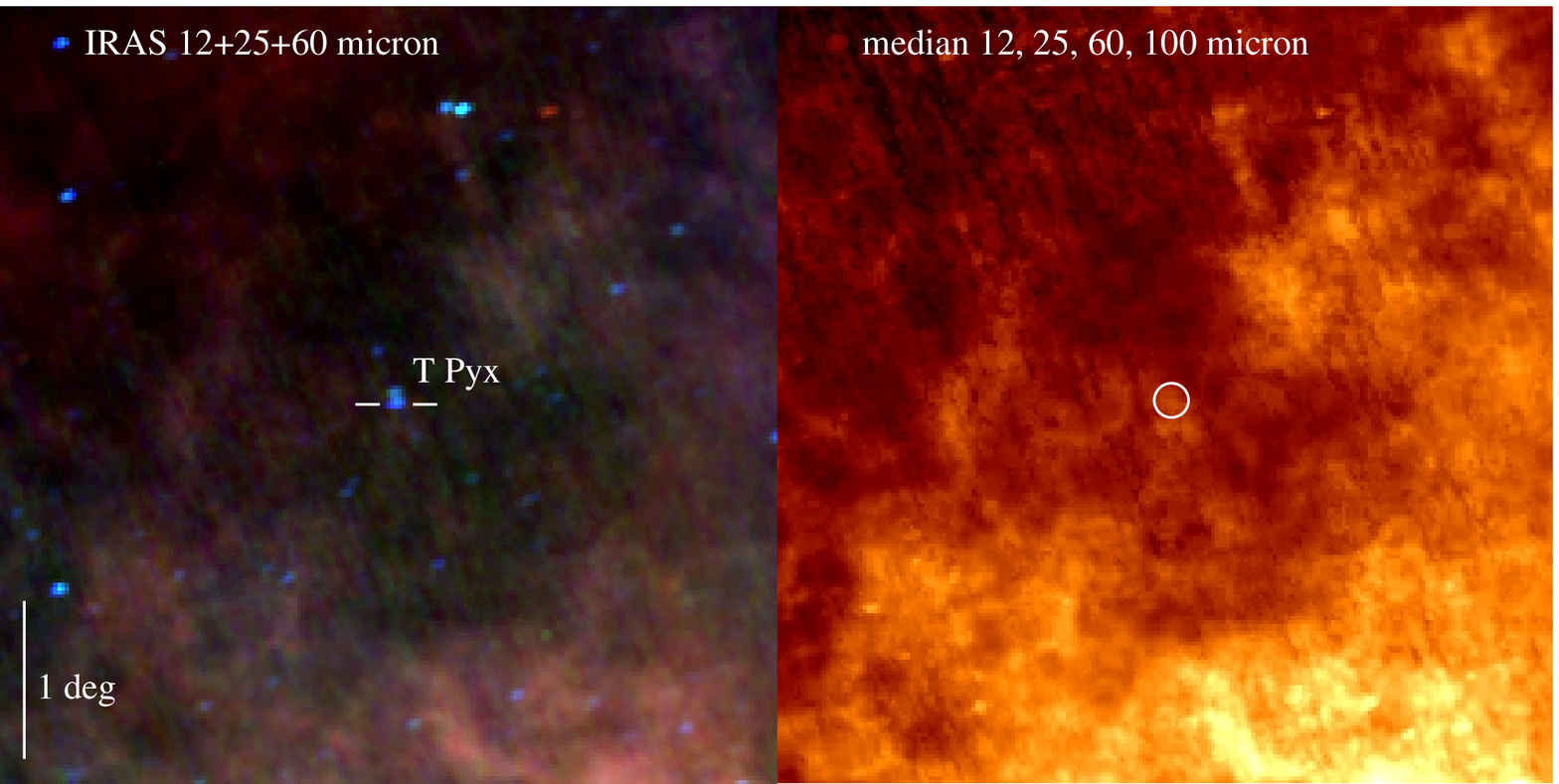}
\caption{IRAS composites around T\,Pyx (North is up, East is left); the right
panel is a median combination of scaled 12, 25, 60 and 100 $\mu$m images.}
\end{figure}

IRAS images around the recurrent dwarf nova T\,Pyxidis \citep[and references
therein]{GilmozziSelvelli2007} also show a cavity, with a $2^{\prime\prime}$
diameter (Fig.\ 9). A $10^{\prime\prime}$ diameter hot shell was linked with
the 1944 eruption or an earlier one by \citet{SharaEtal1989}; the larger
cavity I report here could have originated more than $10^4$ yr ago, possibly
built up during successive eruptions.

\section{Concluding remarks}

Dust in RS\,Oph has been detected in between several eruptions, and at only
half a year since the 2006 eruption. The silicate dust species and mass-loss
rate of $\dot{M}\sim2$ to $3\times10^{-8}$ M$_\odot$ yr$^{-1}$ are not at odds
with the expectations for well-developed dust-driven winds of single stars.
But because RS\,Oph has a relatively warm photosphere, weak pulsation and not
very high luminosity, this might in fact suggest that the mass-loss rate {\it
is} enhanced by the effects of the companion white dwarf. The eruptions do
{\it not} seem to have a dramatic effect on the dust envelope. An accretion
rate onto the white dwarf surface of $\dot{M}>10^{-7}$ M$_\odot$ yr$^{-1}$
would probably require an enhanced flow through the Lagrangian point L$_1$.

Considering the pattern in the time intervals between historic eruptions,
including the 1907 eruption \citep{Schaefer2004}, one might heuristically
expect the next eruption to be due around 2015 (Fig.\ 10).

\begin{figure}[!b]
\plotone{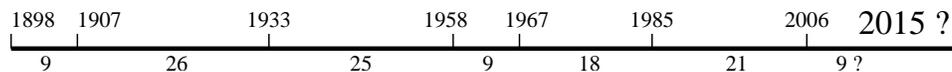}
\caption{Recent outbursts of RS\,Oph and a heuristic prediction for the next.}
\end{figure}

\acknowledgements 
I would like to thank everyone for a pleasant meeting, and Joana Oliveira for
helping me, specifically with making the colour images.

\end{document}